\documentclass[twoside, reqno]{amsart}
\newcommand{\Inline}[1]{\small #1 \normalsize}

\newtheorem{theorem}{Theorem}[section]
\newtheorem{lemma}[theorem]{Lemma}

\theoremstyle{definition}
\newtheorem{definition}[theorem]{Definition}
\newtheorem{example}[theorem]{Example}

\theoremstyle{remark}

\numberwithin{equation}{section}

\begin{document}

\title[Comparative Computational Strength of Quantum Oracles]{COMPARATIVE COMPUTATIONAL STRENGTH OF\\ QUANTUM ORACLES}

\author{Alp At\i c{\i}}

\address{Department of Mathematics, Columbia University, 
2990 Broadway, New York, NY 10027}
\email{atici@math.columbia.edu}



\begin{abstract}
  It is an established fact that for many of the interesting problems quantum algorithms based on queries 
  of the standard oracle bring no significant improvement in comparison to known classical algorithms.
  It is conceivable that there are other oracles of algorithmic importance acting in a less intuitive 
  fashion to which such limitations do not apply. Thus motivated this article suggests a broader 
  understanding towards what a general quantum oracle is.

  We propose a general definition of a quantum oracle and give a classification of quantum oracles based on 
  the behavior of the eigenvalues and eigenvectors of their queries. Our aim is to determine the computational 
  characteristics of a quantum oracle in terms of these eigenvalues and eigenvectors. Within this framework we 
  attempt to describe the class of quantum oracles that are efficiently simulated by the standard oracle and 
  compare the computational strength of different kinds of quantum oracles by an adversary argument using 
  trigonometric polynomials. 
 \end{abstract}

\maketitle
\section{Motivation}
An important class of quantum algorithms makes use of ``quantum oracles'', a device due to which the quantum
circuit solving the problem becomes a function of part of the input. For such algorithms the number of invocations 
of quantum oracles is crucial and defines the query complexity of the problem. The most basic quantum oracle is 
the standard oracle whose computational strength is investigated extensively (See \cite{BBCMdW, BBBV}). 

The important result of \cite{BBCMdW} is that one could represent the coefficients of elements of the computational 
basis with polynomials of input variables when queries of the standard oracle are considered. By an argument on 
the increase in the degree of those polynomials appearing as coefficients for successive queries of the standard oracle 
combined with another argument on the minimum degree of an approximating polynomial required to yield the desired 
function, one could calculate a lower bound on the necessary number of queries determining the query complexity. 
Unfortunately the applicability of this idea is limited to the standard oracle. 

Rather than arguing on the coefficients of the computational basis \cite{FGGS} argues on the basis formed by the 
eigenvectors of the standard oracle queries to calculate such a lower bound. \cite{BESSEN} makes use of an analytic 
argument based on the degrees considering the coefficients of elements of the computational basis as trigonometric 
polynomials to obtain a result comparing the standard oracle to the phase oracle. 

The approach in this article uses trigonometric polynomials of \emph{eigenvalues of an oracle query} to represent 
the coefficients of states obtained by the quantum circuit to generalize the applicability of the argument used 
to compare two oracles in \cite{BESSEN}. To be able to apply an adversary argument to compare a larger class of 
oracles we start with a generalized definition of a quantum oracle and propose a classification of quantum oracles 
based on the behavior of eigenvalues and eigenvectors of an oracle query. This classification will ultimately 
provide the required framework for the comparison of a pair of quantum oracles in a general setting. 

Such a general approach towards quantum oracles is significant because it is closer to the circuit description 
model than the usual black-box model, any block used repetitively in some algorithm could be regarded as an oracle query. 
Many problems such as the graph isomorphism problem are likely to give way to an efficient solution in the presence 
of a particular quantum oracle. Thus a classification of relative strength of a large class of quantum oracles 
could provide useful insight especially when coupled with the knowledge of the class of functions at which a  
quantum oracle could be constructed by a polynomial network of unitary gates.

\section{Quantum Oracles and Quantum Algorithms with Oracle Que\-ries}

\subsection{Introduction and definitions}

Let $\mathbb{B}=\{0,1\}$ and $\mathbb{F}_{n}^{m}$ be the class of functions of the form 
$f: \mathbb{B}^{n}\rightarrow \mathbb{B}^{m}$. The class of all such discrete valued functions 
defined over a discrete set is given by ${\displaystyle\mathbb{F}=\cup_{n, m>0} \mathbb{F}_{n}^{m}}$.
Let the set of all unitary transformations given by a matrix of size $2^{k}\times 2^{k}$ be 
$\mathfrak{U}_{k}$ and $\mathfrak{U}=\cup_{k} \mathfrak{U}_{k}$. It is assumed we are working on a 
quantum system of size $M$ which is sufficiently large for the quantum circuits to operate on.

In the quantum computation literature when an oracle query is mentioned, it is generally implicit
that this is a query of the standard oracle associated to $f \in \mathbb{F}_{n}^{m}$ of the form as
in Section~\ref{bitq}. For quantum algorithms with standard oracle queries, lower bounds for the number 
of queries required to calculate typical functions are established. Calculating \textsc{Or} and 
\textsc{Parity} of $f \in \mathbb{F}_{n}^{1}$ has query complexities $\Theta(\sqrt{2^{n}})$ 
and $\Theta(2^{n})$ respectively (See \cite{BBCMdW,FGGS}). These results are often 
attributed to an intrinsic weakness of the quantum computation model involving queries 
(See \cite{NICHU} pp.~271).  Our conviction is that the 
definition of the standard oracle is far too ``classical'' to bring out the full potential of a quantum 
system and thus mainly responsible for such unsavory results. It is likely that there are other oracles 
of algorithmic importance encoding the information in a less trivial fashion some of which can be 
constructed with a polynomial network at those functions the standard oracle could not. Therefore both 
in an attempt to be free from the limitations the standard oracle imposes and to generalize the idea 
of a quantum oracle we'll take a different approach and suggest the following definitions. 

\begin{definition}[Quantum oracles] 
  A \emph{quantum oracle} $Q$ is a collection of functions\\ 
  $\{Q(n,m): \mathbb{F}_{n}^{m}\rightarrow\mathfrak{U}\ |\ n, m \in \mathbb{Z}^{+}\}$.
  The class of all quantum oracles is denoted by $\mathfrak{Q}$. 
\end{definition}

Thus an oracle $Q$ gives rise to a unique unitary linear transformation $Q(n,m)(f)$ when queried at 
$f \in \mathbb{F}_{n}^{m}$. The functions $Q(n,m)$ are called \emph{the instances} of $Q$. With an abuse 
of notation $Q(n,m)(f)$ is denoted by $Q_{f}$. Therefore an oracle query corresponds to a function 
$Q: \mathbb{F}\rightarrow\mathfrak{U}$. It is assumed that one can also \emph{inverse query}
an oracle $Q$ at $f$ to obtain the unitary linear transformation $(Q_{f})^{-1}$.

\begin{definition}[Quantum algorithm with oracle queries] 
  Let $F: \mathbb{F}_{n}^{m}\times\mathbb{B}^{k}\rightarrow\mathbb{B}^{l}$. Upon input $f\in\mathbb{F}_{n}^{m}$ and $x\in\mathbb{B}^{k}$,
  the quantum circuit $\mathbf{U}$ with oracle queries computing $F$ is started with the initial state $|\psi\rangle|x\rangle$, 
  is composed of finitely many unitary linear transformations: $U_{i}$, independent of $f$ and $x$, and queries and inverse 
  queries of $Q^{(\ell)}$ at $f$, independent of $x$, and satisfies: 
  \[
  \forall f\in\ \mathbb{F}_{n}^{m}\ \mathrm{and}\ x\in\mathbb{B}^{k}, \sum_{z\in \mathbb{B}^{M-k-l}} {|\langle x|\langle z,F(f,x)|\mathbf{U}|\psi\rangle|x\rangle|}^{2}\geq 1-\epsilon
  \]
  \[
  \forall f\in\ \mathbb{F}_{n}^{m}\ \mathrm{and}\ x\in\mathbb{B}^{k}, \sum_{w\in \mathbb{B}^{M-k}} {|\langle x|\langle w|\mathbf{U}|\psi\rangle|x\rangle|}^{2}=1
  \] 
  where $\epsilon$ is a fixed number smaller than $\frac{1}{2}$. Such a quantum circuit $\mathbf{U}$ with 
  queries implements a quantum algorithm with queries computing $F$.
\end{definition}

Note that with this definition the actual input is encoded both in the function
$f$ and the state $|x\rangle$.

\begin{definition}[An oracle approximating another oracle at functions in $\mathbb{F}_{n}^{m}$] Given two oracles $Q^{(1)}$, $Q^{(2)}$ 
  to be queried at the same $f\in\mathbb{F}_{n}^{m}$, if there exists a quantum circuit $\mathbf{U}$ involving $N$ 
  queries (and inverse queries) of $Q^{(1)}$ such that $\|Q^{(2)} - \mathbf{U}\|<\epsilon$    
  independently of $f\in\mathbb{F}_{n}^{m}$ then $Q^{(1)}$ is said to approximate $Q^{(2)}$ up to $\epsilon$ with $N$ queries 
  at functions in $\mathbb{F}_{n}^{m}$. If there exists such an $N$ for which $\epsilon$ could be taken arbitrarily small, then 
  $Q^{(1)}$ is said to simulate $Q^{(2)}$ with $N$ queries at functions in $\mathbb{F}_{n}^{m}$. 
\end{definition}

\begin{definition}[Efficient approximation and simulation between oracles]
	For arbitrary oracles, $Q^{(1)}$ is said to \emph{efficiently approximate} $Q^{(2)}$ up to $\delta$ if it 
	could approximate $Q^{(2)}$ up to $\delta$ with $O(poly(n+m))$ queries at functions in $\mathbb{F}_{n}^{m}$ 
	for any choice of $n$ and $m$. Efficient simulation is defined analogously.
\end{definition}

If $Q^{(1)}$ can approximate $Q^{(2)}$ efficiently but not the other way around this could imply existence of 
a function whose query complexity would be lower if queries of $Q^{(1)}$ could be employed. Moreover $Q^{(1)}$ 
would be at least as powerful as $Q^{(2)}$ for any purpose of computation.
\subsection{Examples of quantum oracles}

\subsubsection{The standard oracle}
\label{bitq}
The standard oracle $Q^{std}$ is the oracle whose query at $f \in \mathbb{F}_{n}^{m}$ gives rise to
$Q_{f}^{std}\in\mathfrak{U}_{n+m}$ with the following action on the computational basis of a quantum system 
of $n+m$ qubits:
\begin{equation}
  Q_{f}^{std}: |x\rangle|y\rangle\mapsto |x\rangle|y\oplus f(x)\rangle
\end{equation}
where $\oplus$ denotes summation modulo $2^{m}$.

For $0\leq s \leq 2^{m}-1$ let 
\begin{equation}
  |\psi_{s}\rangle = \frac{1}{2^{m/2}} \sum_{k=0}^{2^{m}-1} e^{\frac{-2\pi \mathit{i} s k}{2^{m}}} |k\rangle
\end{equation}

Then it is easy to verify for each $0\leq x \leq 2^{n}-1$ and $0\leq s \leq 2^{m}-1$
\begin{align}
  Q_{f}^{std} (|x\rangle|\psi_{s}\rangle) &= |x\rangle \otimes (\frac{1}{2^{m/2}} \sum_{k=0}^{2^{m}-1} e^{\frac{-2\pi \mathit{i} s k}{2^{m}}} |k\oplus f(x)\rangle)\notag\\ 
  &=|x\rangle \otimes (\frac{1}{2^{m/2}} \sum_{k=0}^{2^{m}-1} e^{\frac{-2\pi \mathit{i} s (k-f(x))}{2^{m}}} |k\rangle)\notag\\ 
  &=e^{\frac{2\pi \mathit{i} s f(x)}{2^{m}}} |x\rangle|\psi_{s}\rangle
\end{align}

\subsubsection{The complex phase oracle}
The complex phase oracle $Q^{cp_{d}}$ of degree $d$ is the oracle whose query at $f \in \mathbb{F}_{n}^{m}$ gives rise to
$Q_{f}^{cp_{d}}\in\mathfrak{U}_{n}$ with the following action on the computational basis of a quantum system 
of $n$ qubits:
\begin{equation}
  Q_{f}^{cp_{d}}: |x\rangle \mapsto  e^{\frac{2\pi\mathit{i} d f(x)}{2^{m}}} |x\rangle
\end{equation}

Therefore the query $Q_{f}^{cp_{d}}$ encodes the information purely into the phase of the eigenvalue associated to each computational basis state. 

\subsubsection{The minimal oracle}
The minimal oracle $Q^{min}$ is the oracle whose query at $f \in \mathbb{F}_{n}^{m}$ yields the identity operator in 
$\mathfrak{U}_{n}$ unless $n=m$ and $f$ is a permutation on the set $\mathbb{B}^{n}$. In that case a query of 
$Q^{min}$ at $f$ gives rise to $Q_{f}^{min}\in\mathfrak{U}_{n}$ with the following action on the computational 
basis of a quantum system of $n$ qubits:
\begin{equation}
  Q_{f}^{min}: |x\rangle \mapsto |f(x)\rangle 
\end{equation}

The eigenvectors of such an oracle query depends on the orbits of the permutation $f$. 
Let the permutation $f\in \mathbb{F}_{n}^{n}$ have $\mathfrak{o}$ orbits, $r_{\ell}$ be the length of the 
$\ell$-th orbit and $x_{\ell}$ be the smallest element of the $\ell$-th orbit. 
For all $1\leq \ell \leq \mathfrak{o}$ and $0\leq s_{\ell} \leq r_{\ell}-1$ the following vectors are easily
observed to be the eigenvectors:
\begin{equation}
  |\psi_{{\ell},s_{\ell}}\rangle = \frac{1}{\sqrt{r_{\ell}}} \sum_{k=0}^{r_{\ell}-1} e^{\frac{-2\pi \mathit{i} s_{\ell} k}{r_{\ell}}} |f^{k}(x_{\ell})\rangle\quad \mathrm{associated\ to\ eigenvalue}\ e^{\frac{2\pi \mathit{i} s_{\ell}}{r_{\ell}}}
\end{equation}

The minimal oracle has immense algorithmic importance. As \cite{KKVB} noted, the minimal oracle 
could give an efficient solution to a restricted version of a graph isomorphism problem. 
Order finding algorithm of \cite{SHOR} could also be considered to employ minimal oracle 
queries.

It is a known result that although $Q^{min}$ could simulate $Q^{std}$ with two queries at permutations in 
$\mathbb{F}_{n}^{n}$, to simulate $Q^{min}$ via $Q^{std}$ at permutations in $\mathbb{F}_{n}^{n}$, 
$O(\sqrt{2^n})$ oracle queries are necessary (See \cite{KKVB}).
\section{A Classification of Quantum Oracles}
\label{sec:cqo}
We shall present a classification of oracles based on the eigenvectors and eigenvalues of their queries.

Assume we are provided with an oracle $Q$. Each instance $Q(n,m)$ could be lifted so that 
\begin{equation}
  Q(n,m) : \mathbb{F}_{n}^{m}\rightarrow\mathfrak{U}_{r}
\end{equation} for some sufficiently large $r$ by redefining 
$Q(n,m)(f)$ to be $(Q(n,m)(f)\otimes I) \in \mathfrak{U}_{r}$ whenever $Q(n,m)(f) \in \mathfrak{U}_{s}$ and 
$s<r$ originally. Therefore each instance of the oracle $Q$ is assumed to be in this form where $r\geq n$ 
is the smallest such integer to allow such a lift.

The instance $Q(n,m): \mathbb{F}_{n}^{m}\rightarrow\mathfrak{U}_{r}$ with $r=n+k$ is called 
\emph{nonentangled} if at any choice of $f\in\mathbb{F}_{n}^{m}$ there exists an orthonormal basis 
$\{|\alpha_{x}\rangle,\ 0\leq x \leq 2^{n}-1\}$ for ${(\mathbb C^{2})}^{\otimes n}$ such that 
$Q(n,m)(f)$ acts stably on the following subspaces: 
\begin{equation}
  W_{x} = \langle |\alpha_{x}\rangle|0\rangle, |\alpha_{x}\rangle|1\rangle,\ldots ,|\alpha_{x}\rangle|2^{k}-1\rangle \rangle
\end{equation} 
In other words it is required there exist such subspaces $W_{x}$ possibly depending on the choice of 
$f\in\mathbb{F}_{n}^{m}$ and satisfying $Q_{f} (W_{x})=W_{x}$. 

Given such a nonentangled instance $Q(n,m)$, let the pair $(x,i)$ take values in 
$\{0\ldots2^{n}-1\}\times\{0\ldots2^{k}-1\}$. At any choice of $f\in\mathbb{F}_{n}^{m}$ let $|\beta_{x,i}\rangle$ 
be the vectors in ${(\mathbb C^{2})}^{\otimes k}$ such that for each $(x,i)$
\begin{equation}
Q(n,m)(f) (|\alpha_{x}\rangle|\beta_{x,i}\rangle) = e^{\mathit{i}\theta_{x,i}(f)} |\alpha_{x}\rangle|\beta_{x,i}\rangle
\end{equation}
Then the state $|\psi_{x,i}\rangle=|\alpha_{x}\rangle|\beta_{x,i}\rangle$ is called the $(x,i)$-th eigenvector of $Q_{f}$.

The nonentangled instance $Q(n,m)$ is called \emph{basic} if for each $(x,i)$ there exists a real-valued function 
$g_{x,i}$ independent of the choice of $f\in\mathbb{F}_{n}^{m}$ such that after a renumbering of eigenvectors 
if necessary, the phase of the $(x,i)$-th eigenvalue could be expressed entirely as a function of 
$f(x)$: $\theta_{x,i}(f)=g_{x,i}(f(x))$. This definition does not exclude the possibility of the eigenvectors of 
$Q(n,m)(f)$ being dependent on $f$. If the eigenvectors of a basic instance $Q(n,m)$ could further be chosen 
independently of $f\in\mathbb{F}_{n}^{m}$ such an instance is called \emph{simple}.

An oracle is called to be of any of these particular types if all its instances are so up to a base change.
All the previous examples of oracles are nonentangled. The standard oracle and the complex phase
oracle are simple. For the minimal oracle, one may renumber the eigenvectors identifying the index 
$({{\ell},s_{\ell}})$ with $f^{s_{\ell}}(x_{\ell})$ and observe it is not basic because the knowledge of 
$f(x)$ is not sufficient to determine the length of the orbit containing $x$ and any other renumbering
would suffer from a similar problem.

Many quantum algorithms with que\-ries involve estimating one phase $\theta_{x,i}(f)$ using which 
calculation of $F(f,x)$ could be performed efficiently on a classical computer.

\section{On the Class of Nonentangled Oracles Efficiently Simulated by the Standard Oracle}
\begin{definition}[Locally basic oracles] A nonentangled oracle $Q$ is called \emph{locally basic}, if 
  at any instance $(n,m)$ there exists a function $G\in\mathbb{F}$ and a quantum circuit $\mathbf{U}$ 
  consisting of $O(poly(n+m))$ standard oracle queries satisfying 
  $\mathbf{U}: |\psi_{x,i}\rangle|0\ldots0\rangle \mapsto |\psi_{x,i}\rangle|G(f,x,i)\rangle$ 
  where $|\psi_{x,i}\rangle$ is the $(x,i)$-th eigenvector of a query of $Q$, such that each 
  $\theta_{x,i}(f)$, the phase of its $(x,i)$-th eigenvalue, is entirely a function of $G(f,x,i)$. 
\end{definition}

Clearly any basic oracle is locally basic. 
Any locally basic oracle could be efficiently simulated by the standard oracle since there exists a 
quantum circuit involving polynomially many standard oracle queries mapping: 
\begin{equation}
	\label{lbo}
	|\psi_{x,i}\rangle|0\ldots0\rangle {\underset{\mathbf{U}}{\mapsto}} |\psi_{x,i}\rangle|G(f,x,i)\rangle \mapsto 
	e^{\mathit{i}\theta_{x,i}(f)} |\psi_{x,i}\rangle|G(f,x,i)\rangle {\underset{{\mathbf{U}}^{-1}}{\mapsto}} 
	e^{\mathit{i}\theta_{x,i}(f)} |\psi_{x,i}\rangle|0\ldots0\rangle \quad \forall (x,i)
\end{equation}

\begin{example}$Q^{std}$ could efficiently simulate $Q^{min}$ at permutations in $\mathbb{F}_{n}^{n}$ 
	whose every orbit has $O(poly(n))$ length.
\end{example}

Consider the set of permutations in the classes of functions $\mathbb{F}_{n}^{n}$ whose every orbit has 
length at most $p(n)$, for some polynomial $p(x)$. At such permutations, the minimal oracle behaves like
a locally basic oracle. Consider the following quantum circuit $\mathbf{V}$:
\begin{enumerate}
  \item Apply $Q^{std}$ $p(n)$ times to obtain $|x\rangle|0\ldots0\rangle \mapsto |x\rangle|f(x)\rangle|f^{2}(x)\rangle\ldots|f^{p(n)}(x)\rangle|0\ldots0\rangle$.
  \item By going through the list of elements calculate the length of the orbit (guaranteed to be smaller 
	  than $p(n)$ by assumption), $r$, and the number of iterations of $f$ necessary to reach $x$ starting from
	  the smallest element in the orbit, $s$:\\
	  $|x\rangle|f(x)\rangle|f^{2}(x)\rangle\ldots|f^{p(n)}(x)\rangle|0\ldots0\rangle \mapsto |x\rangle|f(x)\rangle|f^{2}(x)\rangle\ldots|f^{p(n)}(x)\rangle|r\rangle|s\rangle$.
  \item Apply the step $(1)$ in reverse to obtain $|x\rangle|0\ldots0\rangle|r\rangle|s\rangle$.
\end{enumerate}
A similar argument reveals the existence of the quantum circuit $\mathbf{W}$, involving polynomially many queries of 
the standard oracle mapping each $|f^{s_{\ell}}(x_{\ell})\rangle|0\ldots0\rangle$ to 
$|\psi_{{\ell},s_{\ell}}\rangle|0\ldots0\rangle$. Thus the following algorithm involves polynomially many 
queries of the standard oracle:
\begin{align*}
	&|\psi_{{\ell},s_{\ell}}\rangle|0\ldots0\rangle {\underset{{\mathbf{W}}^{-1}}{\mapsto}} |f^{s_{\ell}}(x_{\ell})\rangle|0\ldots0\rangle {\underset{\mathbf{V}}{\mapsto}} |f^{s_{\ell}}(x_{\ell})\rangle|0\ldots0\rangle|r_{\ell}\rangle|s_{\ell}\rangle {\underset{\mathbf{W}}{\mapsto}} \\ 
	&|\psi_{{\ell},s_{\ell}}\rangle|0\ldots0\rangle|r_{\ell}\rangle|s_{\ell}\rangle {\underset{\textrm{swap}}{\mapsto}} |\psi_{{\ell},s_{\ell}}\rangle|r_{\ell}\rangle|s_{\ell}\rangle|0\ldots0\rangle \quad \forall ({\ell},s_{\ell})
\end{align*} 
Because each eigenvalue of a minimal oracle query is entirely a function of $r_{\ell}$ and $s_{\ell}$, one can 
repeat the argument in (\ref{lbo}) to obtain the result.

However due to \cite{KKVB} the standard oracle could not efficiently simulate the minimal oracle at all the
permutations and thus could not be locally basic. This fact is reminiscent of \cite{OZHI} when the 
eigenvalues of the minimal oracle is considered. 

Conversely one might wonder if an oracle $Q$ that is efficiently simulated by the standard oracle has to
be locally basic. Example \ref{fex} comparing the standard oracle with a generic local phase oracle
implies that a direct approach to this question, trying to construct an algorithm with polynomially 
many queries of $Q$ mapping 
$|\psi_{x,i}\rangle|0\ldots0\rangle \mapsto |\psi_{x,i}\rangle|G(f,x,i)\rangle\ \forall (x,i)$, 
is bound to fail for a large class of encodings of $G(f,x,i)$ into the phase of the eigenvalues. 
Thus it is conceivable that locally basic oracles describe a strictly smaller class.

\section{A Comparison of Nonentangled Quantum Oracles}
We shall assume we are given nonentangled oracles $Q^{(1)}, Q^{(2)}$ to compare in terms of efficient 
approximation and therefore consider an arbitrary instance $(n,m)$ of $Q^{(1)}$ and $Q^{(2)}$. 

As earlier assume we are working on a quantum system of sufficiently large size $M$ with an ordering on qubits. 
Without loss of generality we also assume oracle queries are always acting on the first qubits in this order 
by modifying all constant unitary transformations used in oracle approximation to account for the base change. Therefore 
to decide the action of an oracle query at functions in $\mathbb{F}_{n}^{m}$ on this quantum system, we will assume 
the instance $(n,m)$ of both oracles are given in the following form after a lift as in Section~\ref{sec:cqo}: 
\begin{equation}
Q^{(\ell)}: \mathbb{F}_{n}^{m}\rightarrow\mathfrak{U}_{M}\quad \ell=1, 2
\end{equation}
In view of the earlier requirement in Section~\ref{sec:cqo}, $M$ not being minimal does not affect the properties 
satisfied but only changes the multiplicity of eigenvalues. 

Our aim is to put a lower bound on the $Q^{(1)}$ queries required to approximate $Q^{(2)}$ up to $\delta$ at 
functions in $\mathbb{F}_{n}^{m}$ using a classical adversary argument and show that this lower bound grows 
faster than any polynomial in $n+m$. Based on this framework of trigonometric polynomials of the eigenvalues of
a query one may also make use of a quantum adversary argument in the sense of \cite{AMBA} to compare 
nonentangled oracles.

To conclude more than $N$ queries of $Q^{(1)}$ is necessary to approximate $Q^{(2)}$ up to $\delta$ at functions 
in $\mathbb{F}_{n}^{m}$ it is sufficient to show the existence of $f\in\mathbb{F}_{n}^{m}$ and an eigenvector 
$|\psi^{(2)}_{x,i}\rangle$ of $Q_{f}^{(2)}$ for some $(x,i)\in\{0\ldots2^{n}-1\}\times\{0\ldots2^{M-n}-1\}$ such that
\begin{equation}
  \label{eq3}
  \|Q_{f}^{(2)}|\psi^{(2)}_{x,i}\rangle-U_{(N+1)}(Q_{f}^{(1)})^{p_{N}}U_{N}(Q_{f}^{(1)})^{p_{N-1}}\cdots(Q_{f}^{(1)})^{p_{1}}U_{1}|\psi^{(2)}_{x,i}\rangle\|>\delta
\end{equation}
for any choice of $\{U_{k}\in\mathfrak{U}_{M}\ |\ 1\leq k\leq N+1\}$ and $\{p_{k}=\pm 1\ |\ 1\leq k \leq N\}$. 

It is not necessary in general that an eigenvector satisfying (\ref{eq3}) exists even if approximation with $N$ 
queries up to $\delta$ is impossible. However this is necessary for impossibility of simulation with $N$ queries, 
i.e. if for all eigenvectors of $Q_{f}^{(2)}$ the following equation is satisfied:
\begin{equation}
U_{(N+1)}(Q_{f}^{(1)})^{p_{N}}U_{N}(Q_{f}^{(1)})^{p_{N-1}}\cdots(Q_{f}^{(1)})^{p_{1}}U_{1}|\psi^{(2)}_{x,i}\rangle=Q_{f}^{(2)}|\psi^{(2)}_{x,i}\rangle
\end{equation}
for a particular choice of $\{U_{k}\in\mathfrak{U}_{M}\ |\ 1\leq k\leq N+1\}$ and $\{p_{k}=\pm 1\ |\ 1\leq k \leq N\}$
then by linearity the same would clearly hold for any possible initial state. Therefore one could 
immediately deduce that $Q^{(2)}$ could be simulated with at most $N$ queries of $Q^{(1)}$ at 
functions in $\mathbb{F}_{n}^{m}$.

\begin{definition}[Trigonometric polynomial]
	\label{trpoly}
  A function $T: \mathbb{R}^{D}\rightarrow\mathbb{C}$ of the form:
  \[
  T(\phi_{1},\ldots,\phi_{D})=\sum_{j=1}^{K} c_{j} e^{\mathit{i}(n_{j,1}\phi_{1}+\ldots+n_{j,D}\phi_{D})}\ \mathrm{where}\ c_{j}\in\mathbb{C},\ n_{j,k}\in\mathbb{Z}\ \mathrm{and}\ K<\infty 
  \]
  is called an \emph{$D$-variate trigonometric polynomial} with \emph{degree}
  \[
  \mathbf{deg}\ T(\phi_{1},\ldots,\phi_{D})=\max_{j}(|n_{j,1}|+\ldots|n_{j,D}|)
  \]
\end{definition}

The following lemma is a generalized yet simpler variant of the one in \cite{BESSEN} 
but considered in the framework of eigenvectors and eigenvalues. The approach in \cite{BBCMdW}
is based on degree of ordinary polynomials but its applicability is limited to the 
standard oracle.
\begin{lemma}
  \label{thm1}
  Let $Q(n,m)$ be a nonentangled instance with $\theta_{x,i}$ corresponding to the phase of the $(x,i)$-th 
  eigenvalue and $|\psi\rangle$ be the state of the quantum system represented 
  with respect to a particular choice of orthonormal basis $\{|\alpha_{j}\rangle\ |\ 0\leq j \leq 2^{M}-1\}$ 
  as follows:
  \[  
  |\psi\rangle=\sum_{j=0}^{2^{M}-1} T_{j}(\mathbf{\Theta}(f)) |\alpha_{j}\rangle\ \mathrm{where}\ \mathbf{\Theta}(f)=\{\theta_{x,i}(f)\ |\ 0\leq x \leq 2^{n}-1, 0\leq i \leq 2^{M-n}-1\}
  \]
  where each $T_{j}(\phi_{0,0},\ldots,\phi_{x,i},\ldots\phi_{2^{n}-1,2^{M-n}-1})$ is a trigonometric polynomial.
  Let the degree $\mathbf{deg}\ |\psi\rangle=\max_{j} (\mathbf{deg}\ T_{j})$. Then for any unitary operator 
  $U$ independent of $f$, $\mathbf{deg}\ (U|\psi\rangle) \leq \mathbf{deg}\ |\psi\rangle$ and 
  $\mathbf{deg}\ ((Q_{f})^{\pm 1}|\psi\rangle) \leq (\mathbf{deg}\ |\psi\rangle)+1$ for $f\in\mathbb{F}_{n}^{m}$.
\end{lemma}
\begin{proof}
  $\mathbf{deg}\ (U|\psi\rangle) \leq \mathbf{deg}\ |\psi\rangle$: Clear considering how a unitary transformation 
  acts on the coefficients and the definition of trigonometric polynomial. This also implies the definition of 
  the degree does not depend on the choice of basis.

  $\mathbf{deg}\ ((Q_{f})^{\pm 1}|\psi\rangle) \leq (\mathbf{deg}\ |\psi\rangle)+1$: It is clear if 
  $\{|\alpha_{j}\rangle\ |\ 1\leq j \leq 2^{M}\}$ are the eigenvectors of $Q_{f}$. Otherwise use 
  the previous result.
\end{proof}

Note that each trigonometric polynomial corresponding to the coefficient of a basis element following the application of
a number of unitary transformations some of which are oracle queries to a unit vector has to be smaller than or equal to 
$1$ in absolute value.

\begin{lemma}
  \label{lemma1} Given two nonentangled oracles $Q^{(1)}$, $Q^{(2)}$ with 
  $\{U_{k}\in\mathfrak{U}_{M}\ |\ 1\leq k\leq N+1\}$ and $\{p_{k}=\pm 1\ |\ 1\leq k \leq N\}$ then
  \Inline{$U_{(N+1)}(Q_{f}^{(1)})^{p_{N}}U_{N}(Q_{f}^{(1)})^{p_{N-1}}\cdots(Q_{f}^{(1)})^{p_{1}}U_{1}$}
  could approximate $Q_{f}^{(2)}$ at most up to
  \Inline{$\max_{x,i} (|e^{\mathit{i}\theta^{(2)}_{x,i}(f)} - T^{x,i}_{x,i}(\mathbf{\Theta}^{(1)}(f))|)$}
  at a function $f\in\mathbb{F}_{n}^{m}$ where 
  \begin{list}{$\bullet$}{\setlength{\leftmargin}{.2in}
			\setlength{\rightmargin}{.2in}
			\setlength{\itemsep}{0ex}}
	\item $\ $ $e^{\mathit{i}\theta^{(2)}_{x,i}(f)}$ denotes the $(x,i)$-th eigenvalue of the oracle query 
	  $Q_{f}^{(2)}$ corresponding to the eigenvector $|\psi^{(2)}_{x,i}\rangle$, also a function of $f$. 
	\item $\ $ $T^{x,i}_{x^{\prime},i^{\prime}}(\mathbf{\Theta}^{(1)}(f))$ denotes the trigonometric polynomial (of 
	  the phases of the eigenvalues of the oracle query $Q_{f}^{(1)}$) corresponding to the scalar projection of the vector \\
	  \Inline{$U_{(N+1)}(Q_{f}^{(1)})^{p_{N}}U_{N}(Q_{f}^{(1)})^{p_{N-1}}\cdots(Q_{f}^{(1)})^{p_{1}}U_{1}|\psi^{(2)}_{x,i}\rangle$}
	  over $|\psi^{(2)}_{x^{\prime},i^{\prime}}\rangle$.
  \end{list}
  Moreover each $T^{x,i}_{x^{\prime},i^{\prime}}(\mathbf{\Theta}^{(1)}(f))$ has degree at most $N$.
\end{lemma}
\begin{proof}
  As stated, given $\{U_{k}\in\mathfrak{U}_{M}\ |\ 1\leq k\leq N+1\}$ and $\{p_{k}=\pm 1\ |\ 1\leq k \leq N\}$ let
  \begin{equation}U_{(N+1)}(Q_{f}^{(1)})^{p_{N}}U_{N}(Q_{f}^{(1)})^{p_{N-1}}\cdots(Q_{f}^{(1)})^{p_{1}}U_{1}|\psi^{(2)}_{x,i}\rangle=
  \sum_{x^{\prime}=0}^{2^{n}-1}\sum_{i^{\prime}=0}^{2^{M-n}-1} T^{x,i}_{x^{\prime},i^{\prime}}(\mathbf{\Theta}^{(1)}(f)) |\psi^{(2)}_{x^{\prime},i^{\prime}}\rangle
  \end{equation}
  
  If vectors are represented with respect to the basis formed by the eigenvectors of the oracle query 
  $Q_{f}^{(2)}$, observing only the difference in the coefficient of the eigenvector $|\psi^{(2)}_{x,i}\rangle$ gives
  \begin{equation}\|Q_{f}^{(2)}|\psi^{(2)}_{x,i}\rangle-U_{(N+1)}(Q_{f}^{(1)})^{p_{N}}U_{N}(Q_{f}^{(1)})^{p_{N-1}}\cdots(Q_{f}^{(1)})^{p_{1}}U_{1}|\psi^{(2)}_{x,i}\rangle\|\geq
  |e^{\mathit{i}\theta^{(2)}_{x,i}(f)} - T^{x,i}_{x,i}(\mathbf{\Theta}^{(1)}(f))|\end{equation}
  As all eigenvectors $|\psi^{(2)}_{x,i}\rangle$ of $Q_{f}^{(2)}$ are considered, this yields the stated 
  result at every $f\in\mathbb{F}_{n}^{m}$.

  Furthermore as a consequence of Lemma~\ref{thm1} each of the trigonometric polynomials\\ 
  $T^{x,i}_{x^{\prime},i^{\prime}}(\mathbf{\Theta}^{(1)}(f))$ has degree at most $N$.
\end{proof}

Therefore demonstration of (\ref{eq3}) reduces to finding a lower bound for 
\Inline{$|e^{\mathit{i}\theta^{(2)}_{x,i}(f)} - T^{x,i}_{x,i}(\mathbf{\Theta}^{(1)}(f))|$} for some $(x,i)$ 
and $f\in\mathbb{F}_{n}^{m}$. Making use of this fact one could determine the impossibility of efficient 
approximation between two oracles in a much more general sense.
\begin{theorem}\label{mainthm} Given two nonentangled oracles $Q^{(1)}$, $Q^{(2)}$ with 
  $\{U_{k}\in\mathfrak{U}_{M}\ |\ 1\leq k\leq N+1\}$ and $\{p_{k}=\pm 1\ |\ 1\leq k \leq N\}$ then
  \Inline{$U_{(N+1)}(Q^{(1)})^{p_{N}}U_{N}(Q^{(1)})^{p_{N-1}}\cdots(Q^{(1)})^{p_{1}}U_{1}$}
  could approximate $Q^{(2)}$ at functions in $\mathbb{F}_{n}^{m}$ at most up to\Inline{
  \[\frac{1}{2} [\max_{f_{1}, f_{2}\in\mathbb{F}_{n}^{m}} \{\max_{x,i} (||e^{\mathit{i}\theta^{(2)}_{x,i}(f_{1})} - T^{x,i}_{x,i}(\mathbf{\Theta}^{(2)}(f_{2}))| - |T^{x,i}_{x,i}(\mathbf{\Theta}^{(1)}(f_{1})) - T^{x,i}_{x,i}(\mathbf{\Theta}^{(1)}(f_{2}))|| ) \} ]\]}
  where 
  \begin{list}{$\bullet$}{\setlength{\leftmargin}{.2in}
			\setlength{\rightmargin}{.2in}
			\setlength{\itemsep}{0ex}}
	\item $\ $ $e^{\mathit{i}\theta^{(2)}_{x,i}(f_{\dot{\jmath}})}$ denotes the $(x,i)$-th eigenvalue of the oracle query 
	  $Q_{f_{\dot{\jmath}}}^{(2)}$.
	\item $\ $ $T^{x,i}_{x^{\prime},i^{\prime}}(\mathbf{\Theta}^{(1)}(f_{\dot{\jmath}}))$ denotes the trigonometric 
	  polynomial (of the phases of the eigenvalues of the oracle query $Q_{f_{\dot{\jmath}}}^{(1)}$) corresponding to 
	  the scalar projection of the vector \\
	  \Inline{$U_{(N+1)}(Q_{f_{\dot{\jmath}}}^{(1)})^{p_{N}}U_{N}(Q_{f_{\dot{\jmath}}}^{(1)})^{p_{N-1}}\cdots(Q_{f_{\dot{\jmath}}}^{(1)})^{p_{1}}U_{1}|\psi^{(2)}_{x,i}\rangle$}
	  over $|\psi^{(2)}_{x^{\prime},i^{\prime}}\rangle$, the $(x^{\prime},i^{\prime})$-th eigenvector of $Q_{f_{1}}^{(2)}$, 
	  therefore has degree at most $N$.
	\item $\ $ $T^{x,i}_{x^{\prime},i^{\prime}}(\mathbf{\Theta}^{(2)}(f_{\dot{\jmath}}))$ denotes the trigonometric 
	  polynomial (of the phases of the eigenvalues of the oracle query $Q_{f_{\dot{\jmath}}}^{(2)}$) corresponding to 
	  the scalar projection of the vector $Q_{f_{\dot{\jmath}}}^{(2)} |\psi^{(2)}_{x,i}\rangle$ over 
	  $|\psi^{(2)}_{x^{\prime},i^{\prime}}\rangle$, the $(x^{\prime},i^{\prime})$-th eigenvector of $Q_{f_{1}}^{(2)}$, 
	  therefore has degree at most $1$.
  \end{list}
  Moreover if $Q^{(2)}$ is a simple oracle then $T^{x,i}_{x,i}(\mathbf{\Theta}^{(2)}(f_{2}))=e^{\mathit{i}\theta^{(2)}_{x,i}(f_{2})}$.
\end{theorem}

\begin{proof}
  Assume the contrary and consider $f_{1}, f_{2}\in\mathbb{F}_{n}^{m}$ and $(x,i)$ for which 
  \begin{equation}
	2 \delta=(||e^{\mathit{i}\theta^{(2)}_{x,i}(f_{1})} - T^{x,i}_{x,i}(\mathbf{\Theta}^{(2)}(f_{2}))| - |T^{x,i}_{x,i}(\mathbf{\Theta}^{(1)}(f_{1})) - T^{x,i}_{x,i}(\mathbf{\Theta}^{(1)}(f_{2}))||)
  \end{equation} is maximized.

  The Triangle Inequality gives:
  \begin{align}
	&|e^{\mathit{i}\theta^{(2)}_{x,i}(f_{1})} - T^{x,i}_{x,i}(\mathbf{\Theta}^{(2)}(f_{2})) + T^{x,i}_{x,i}(\mathbf{\Theta}^{(1)}(f_{2})) - T^{x,i}_{x,i}(\mathbf{\Theta}^{(1)}(f_{1}))|\geq\notag\\
	&||e^{\mathit{i}\theta^{(2)}_{x,i}(f_{1})} - T^{x,i}_{x,i}(\mathbf{\Theta}^{(2)}(f_{2}))| - |T^{x,i}_{x,i}(\mathbf{\Theta}^{(1)}(f_{2})) - T^{x,i}_{x,i}(\mathbf{\Theta}^{(1)}(f_{1}))||=2 \delta
  \end{align} 

  By original assumption \Inline{$|T^{x,i}_{x,i}(\mathbf{\Theta}^{(2)}(f_{2})) - T^{x,i}_{x,i}(\mathbf{\Theta}^{(1)}(f_{2}))| \leq \delta$},
  otherwise we get a contradiction due to the same argument in Lemma~\ref{lemma1}, because the error in approximation has to be at
  least as large as the difference in the coefficient of $|\psi^{(2)}_{x,i}\rangle$.
  Thus another application of the Triangle Inequality gives:
  \footnotesize
  \begin{align}
	&|e^{\mathit{i}\theta^{(2)}_{x,i}(f_{1})} - T^{x,i}_{x,i}(\mathbf{\Theta}^{(1)}(f_{1}))|= \notag\\
	&|e^{\mathit{i}\theta^{(2)}_{x,i}(f_{1})} - T^{x,i}_{x,i}(\mathbf{\Theta}^{(2)}(f_{2})) + T^{x,i}_{x,i}(\mathbf{\Theta}^{(2)}(f_{2})) - T^{x,i}_{x,i}(\mathbf{\Theta}^{(1)}(f_{2})) + T^{x,i}_{x,i}(\mathbf{\Theta}^{(1)}(f_{2})) - T^{x,i}_{x,i}(\mathbf{\Theta}^{(1)}(f_{1}))|\geq \notag\\
	&||e^{\mathit{i}\theta^{(2)}_{x,i}(f_{1})} - T^{x,i}_{x,i}(\mathbf{\Theta}^{(2)}(f_{2})) + T^{x,i}_{x,i}(\mathbf{\Theta}^{(1)}(f_{2})) - T^{x,i}_{x,i}(\mathbf{\Theta}^{(1)}(f_{1}))| - |T^{x,i}_{x,i}(\mathbf{\Theta}^{(2)}(f_{2})) - T^{x,i}_{x,i}(\mathbf{\Theta}^{(1)}(f_{2}))||\notag\\
	&\geq |2\delta-\delta|=\delta 
  \end{align}
  \normalsize
  which gives rise to another contradiction due to Lemma~\ref{lemma1}. Therefore the error should be at
  least as large as $\delta$ at either $f_{1}$ or $f_{2}$ which implies the result.
  Moreover the degree of the trigonometric polynomials should be as stated due to Lemma~\ref{thm1}.
  
  By definition of the simple oracle, if $Q^{(2)}$ is a simple oracle then the eigenvectors of $Q_{f}^{(2)}$
  will be independent of $f$. Therefore $T^{x,i}_{x,i}(\mathbf{\Theta}^{(2)}(f_{2}))=e^{\mathit{i}\theta^{(2)}_{x,i}(f_{2})}$.
  \end{proof}

Thus the argument is based on the fact that the large amount of change in the eigenvalues of $Q_{f}^{(2)}$ 
as $f\in \mathbb{F}_{n}^{m}$ is altered could not be accommodated by the relatively small change in the trigonometric 
polynomial which is a function of the phases of the eigenvalues of $Q_{f}^{(1)}$ since the change in these eigenvalues 
is also small. This imposes a lower bound on the number of $Q_{f}^{(1)}$ oracle queries we should perform to approximate 
$Q_{f}^{(2)}$ at functions in $\mathbb{F}_{n}^{m}$.

The following lemma, also made use of by \cite{BESSEN}, is helpful 
considered in conjunction with the Theorem~\ref{mainthm}.

\begin{lemma} 
  \label{lemma2}
  Let $T(\theta)$ be a trigonometric polynomial of a single variable $\theta\in\mathbb{R}$ satisfying 
  $|T(\theta)|\leq 1$ for any value of $\theta$. If $|T(\theta_{1}) - T(\theta_{2})|=\lambda$ then 
  $\frac{\lambda}{|\theta_{1}-\theta_{2}|} \leq (\mathbf{deg}\ T(\theta))$.
\end{lemma}

\begin{proof}
The Mean Value Theorem yields:
\begin{equation}
\lambda=|T(\theta_{1}) - T(\theta_{2})|=|T^{\prime}(\theta^{\ast})||\theta_{1}-\theta_{2}|\ \mathrm{for\ some}\ \theta^{\ast}\in [\theta_{1},\theta_{2}]\ \mathrm{assuming}\ \theta_{1} < \theta_{2}
\end{equation}
Combining with Bernstein's Inequality (See \cite{CHENEY})
one obtains:
\begin{equation}
  |T^{\prime}(\theta^{\ast})| \leq (\mathbf{deg}\ T(\theta)) \sup_{-\pi\leq\theta\leq\pi} |T(\theta)|\notag\\
  \leq \mathbf{deg}\ T(\theta)
\end{equation}
\end{proof}

\subsection{Example: the comparison of standard oracle vs. a generic local phase oracle}
\label{fex}
Consider a generic local phase oracle $Q^{(1)}$ whose instance $(n,m)$ has the following action on the 
computational basis of a quantum system of $n$ qubits: 
\Inline{$|x\rangle \mapsto e^{\mathit{i} \sum_{y} c_{x,y} (g(\frac{f(y)}{2^{m}}))}|x\rangle$} 
where $g$ is a real valued differentiable function on $[0,1]$ with $\sup_{t\in[0,1]} |g^{\prime}(t)|\leq B$, 
each $c_{x,y} \in \mathbb{R}$, $\max_{x,y} |c_{x,y}| \leq C$ for some constant $C$ and 
$c_{x,y}=0\ \mathrm{whenever}\ 2^{n}-p(n) > |y-x| >p(n)$ for some polynomial $p(x)$ 
($g$, $C$ and $p$ are independent of the instance $(n,m)$). 
Any such oracle is locally basic and thus is efficiently simulated by the standard oracle.

Set $Q^{(2)}=Q^{std}$ for consistency with the earlier notation. Assume $Q^{(1)}$ can approximate 
$Q^{(2)}$ up to $\delta$ with $N$ queries at functions in $\mathbb{F}_{n}^{m}$.

Let $f_{1}, f_{2} \in \mathbb{F}_{n}^{m}$ given by:
\begin{equation} 
  f_{1}(x)= \begin{cases} 
  2^{m-1} & \textrm{if $x=0$} \\
  0 & \textrm{if $x\neq 0$} 
  \end{cases}\ \qquad 
  f_{2}(x)= \begin{cases} 
  2^{m-1}+1 & \textrm{if $x=0$}\\ 
  0 & \textrm{if $x\neq 0$} 
  \end{cases}  
\end{equation} 

Set the initial state of the quantum system of size $M$ to be 
\begin{equation}
|0^{n}\rangle\otimes(\frac{1}{2^{m/2}} \sum_{k=0}^{2^{m}-1} e^{\frac{-2\pi \mathit{i} 2^{m-1} k}{2^{m}}}|k\rangle)\otimes|0^{M-n-m}\rangle=|0^{n}\rangle|\psi_{2^{m-1}}\rangle|0^{M-n-m}\rangle
\end{equation}
with the same notation as in Section~\ref{bitq}. 

An application of the specific form of Theorem~\ref{mainthm} for the simple oracle $Q^{(2)}$, considering the 
action of the query $Q_{f}^{(2)}\otimes I$ on this initial state together with the earlier eigenvalue 
calculation yields: 
\begin{align} 
  |e^{\mathit{i}\theta^{(2)}_{0,2^{m-1}}(f_{1})} - e^{\mathit{i}\theta^{(2)}_{0,2^{m-1}}(f_{2})}|&=|e^{\frac{2\pi\mathit{i}2^{m-1}2^{m-1}}{2^m}}-e^{\frac{2\pi\mathit{i}2^{m-1}(2^{m-1}+1)}{2^m}}|\notag\\
  &=|e^{\pi\mathit{i}2^{m-1}}-e^{\pi\mathit{i}(2^{m-1}+1)}|=|1-e^{\pi\mathit{i}}|=2
\end{align}
Thus in view of Theorem~\ref{mainthm} we obtain:
\begin{align}
  2\delta&\geq |2-|T^{0,2^{m-1},0}_{0,2^{m-1},0}(\mathbf{\Theta}^{(1)}(f_{1})) - T^{0,2^{m-1},0}_{0,2^{m-1},0}(\mathbf{\Theta}^{(1)}(f_{2}))||\\
  2-2\delta&\leq |T^{0,2^{m-1},0}_{0,2^{m-1},0}(\mathbf{\Theta}^{(1)}(f_{1})) - T^{0,2^{m-1},0}_{0,2^{m-1},0}(\mathbf{\Theta}^{(1)}(f_{2}))|
\end{align}

\Inline{$T^{0,2^{m-1},0}_{0,2^{m-1},0}(\mathbf{\Theta}^{(1)}(f))$} is a trigonometric polynomial with degree at most $N$ in the sense of Definition~\ref{trpoly}. It could also be regarded as a trigonometric polynomial $T(\theta)$ 
of a single variable \Inline{$\theta=g(\frac{f(0)}{2^{m}})$} with degree at most $C\times N$ 
(considering a real number as the degree of this trigonometric polynomial in the obvious sense does not affect the 
applicability of Lemma~\ref{lemma2}). Hence an application of Lemma~\ref{lemma2} on 
$T(\theta)$ by letting \Inline{$\theta_{1}=g(\frac{f_{1}(0)}{2^{m}})$}, \Inline{$\theta_{2}=g(\frac{f_{2}(0)}{2^{m}})$} combined with the Mean Value Theorem yields:
\begin{equation}
	\frac{2-2\delta}{B \frac{1}{2^{m}}}=\frac{2^{m+1}(1-\delta)}{B}\leq\mathbf{deg}\ T(\theta) \leq C\times N
\end{equation}

Thus approximating $Q^{std}$ up to $\delta$ requires at least \Inline{$\lceil\frac{2^{m+1}(1-\delta)}{B.C}\rceil$} 
queries of $Q^{(1)}$ at functions in $\mathbb{F}_{n}^{m}$ and therefore could not be efficient for such a generic
local phase oracle. This result is analogous to that of \cite{BESSEN} but a generic local phase oracle acts 
distinctly from the phase oracle in \cite{BESSEN}.

The following elementary lemma is useful for combining results
regarding efficient simulation and approximation.

\begin{lemma} For arbitrary oracles $Q^{(1)}$, $Q^{(2)}$ and $Q^{(3)}$, if $Q^{(1)}$ can
  efficiently simulate $Q^{(2)}$ and $Q^{(2)}$ can efficiently approximate $Q^{(3)}$ up to $\delta$
  then $Q^{(1)}$ can efficiently approximate $Q^{(3)}$ up to $\delta$. In particular if
  $Q^{(1)}$ can efficiently simulate $Q^{(2)}$ and $Q^{(2)}$ can efficiently simulate $Q^{(3)}$
  then $Q^{(1)}$ can efficiently simulate $Q^{(3)}$.
\end{lemma}

As a consequence to this lemma the obtained result coupled with  
that of \cite{KKVB} yields:

\begin{list}{$\bullet$}{\setlength{\leftmargin}{.2in}
			\setlength{\rightmargin}{.2in}
			\setlength{\itemsep}{0ex}}
  \item $Q^{cp_{d}}$ cannot efficiently simulate $Q^{min}$ for any degree $d$.
  \item $Q^{min}$ can efficiently simulate $Q^{cp_{d}}$ for any $d$ at permutations.
\end{list}

Thus $Q^{min}\succ\hspace{-5pt}\footnote{at permutations.}\ Q^{std}\succ Q^{cp_{d}}$ where ``$\succ$'' denotes efficient simulation in single direction.

\section{Acknowledgements}
\noindent
The author wishes to thank Rocco Servedio, Dave Bayer, Henryk Wo\'{z}niakowski and Joseph Traub 
for their support and fruitful discussions and Jeff Phan for new suggestions and his help in correcting many 
inconsistencies.

\end{document}